\RequirePackage{fix-cm}
\documentclass[twocolumn]{svjour3}
\smartqed
\usepackage{graphicx}
\usepackage{amsmath}
\usepackage{amssymb}
\usepackage{mathtools}
\usepackage{multirow}
\usepackage{rotating}
\journalname{bjp}

\begin{document}

\title{Finite-size scaling considerations on the ground state microcanonical temperature in entropic sampling simulations}
\author{A. A. Caparica \and Cl\'audio J. DaSilva}
\institute{A. A. Caparica \at Instituto de F\'{i}sica, Universidade Federal de Goi\'as, C.P. 131 CEP 74001-970, Goi\^{a}nia, Goi\'{a}s, Brazil \\ \email{caparica@ufg.br}  \and Cl\'audio J. DaSilva \at Instituto Federal de Educa\c{c}\~ao, Ci\^encia e Tecnologia de Goi\'as, CEP 74130-012, Goi\^ania, Goi\'{a}s, Brazil}

\date{Received: \today / Accepted: date}

\maketitle

\begin{abstract}
In this work we discuss the behavior of the microcanonical temperature 
$\frac{\partial S(E)}{\partial E}$ obtained by means of numerical 
entropic sampling studies. It is observed that in almost all cases the slope of 
the logarithm of the density of states $S(E)$ is not infinite in the ground 
state, since as expected it should be directly related to the inverse 
temperature $\frac{1}{T}$. Here we show that these finite slopes are in fact due 
to finite-size effects and we propose an analytic expression $a\ln(bL)$ for the 
behavior of $\frac{\varDelta S}{\varDelta E}$ when $L\rightarrow\infty$. To test 
this idea we use three distinct two-dimensional square lattice models 
presenting second-order phase transitions. We calculated by exact means the 
parameters $a$ and $b$ for the two-states Ising model and for the $q=3$ and $4$ 
states Potts model and compared with the results obtained by entropic sampling 
simulations. We found an excellent agreement between exact and numerical values. 
We argue that this new set of parameters $a$ and $b$ represents an interesting 
novel issue of investigation in entropic sampling studies for different models.
\end{abstract}

\section{Introduction}
\label{intro}

Monte Carlo (MC) entropic sampling 
simulations\cite{Lee93,Oliveira96,Oliveira98,Wang00}  have attracted a great 
deal of attention in the 
last years due to its capability  to calculate directly the entropy $S(E)=k_B\ln 
\Omega(E)$ of a variety of  systems. This definition of entropy in terms of the 
number of microstates $\Omega(E)$ with energy $E$ has been the cornerstone of 
statistical mechanics  since its introduction by Boltzmann\cite{Callen85}. Hence 
the computation of such quantity by numerical means is of great importance to 
accurately locate and characterize phase transitions. The motivation for the 
development of these new  MC methods resides in the fact that conventional 
methods exhibit long time scale  problems. These methods are based on the 
estimation of all possible states (or  configurations) for an energy level $E$ 
of the system which is being studied. Since the partition function for a given 
model in statistical physics can be expressed in terms of the density of states 
$\Omega  (E)$, if one can estimate this quantity with high accuracy it is 
possible to construct the partition function as $Z=\sum_E \Omega(E)e^{-E/k_BT}$, 
essentially  solving the model. Here $k_B$ denotes the Boltzmann constant. 
Therefore, one can  study the full temperature range without further simulation 
runs. The most celebrated approach in this direction is the Wang-Landau (WL) 
sampling \cite{WangLandau01a,WangLandau01b}. This method measures an {\it a 
priori} unknown  density of states of a given system iteratively by performing a 
random walk in energy space and sampling configurations with probability 
proportional to the reciprocal of the density of states, resulting in a ``flat" 
histogram.

Conversely, analyzing the results of any entropic sampling simulation, the 
logarithm of the density of states corresponds exactly to the 
definition of entropy. The Boltzmann constant $k_B$ ensures the agreement with 
the Kelvin scale of temperature, defined by
\begin{equation}\label{temperature}
\frac{1}{T}=\frac{\partial S(E)}{\partial E}.
\end{equation}
One should therefore expect a divergence of $\frac{\partial S(E)}{\partial E}$ 
at the ground state temperature. Nevertheless what we see in 
most plots of the logarithm of the density of states obtained by entropic 
sampling simulations is a clear finite slope in the ground state. The reason for 
this apparent contradiction are the finite-size effects. In fact for finite 
systems, from small to large, the thermodynamics is a rather controversial 
issue\cite{Hill62}. In downscale one have problems defining phase 
transitions\cite{Gulminelli99,Borrmann00}, the equivalence between 
ensembles\cite{Gulminelli03,Fiore13}, the extensivity of thermodynamical 
variables\cite{Gross05} and so on. Accordingly, in this work we intend to shed 
some light on the reason why the slope of the logarithm of the density of 
states, obtained by entropic sampling procedures, is not infinite in the ground 
state, where the inverse temperature diverges. Our proposal may provide a 
supplementary way of investigation for any entropic MC simulation.

The outline of this paper is as follows: In section II we present a finite-size 
scaling analysis of the microcanonical temperature. In 
section III we define the simulation procedure. In section IV the simulation 
results are discussed. Section V is devoted to the summary and concluding 
remarks.

\section{Finite-Size Scaling Behavior of the Microcanonical Temperature}
\label{sec:1}

\begin{figure}[!ht]\centering
\begin{center}
 \includegraphics[width=.7\linewidth,angle=-90]{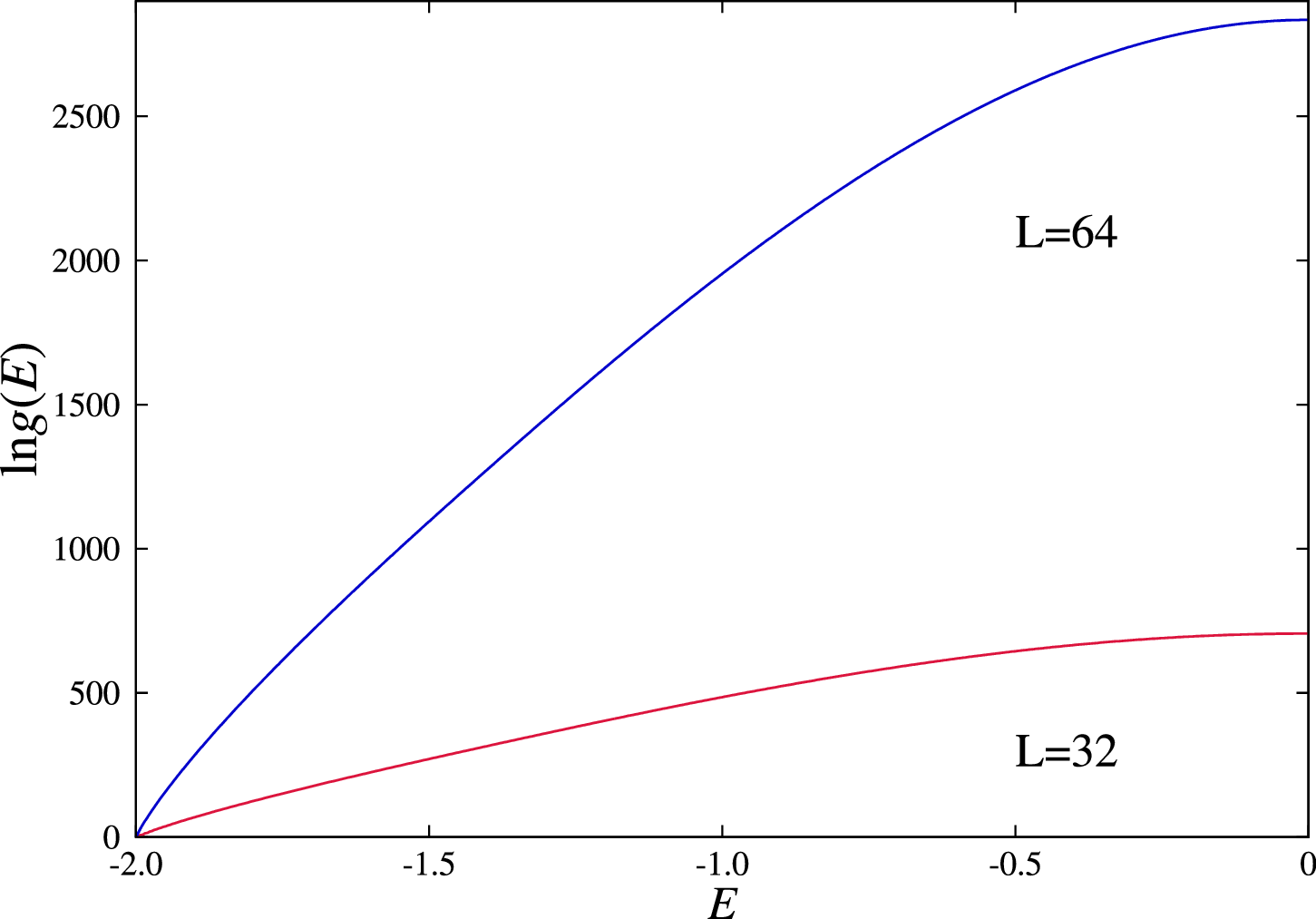}
\end{center}
\caption{(Color online) Logarithm of the density of states obtained by 
Wang-Landau sampling for the 2D Ising model for $L=32,64$ as
functions of the energy per spin.}
\label{fig1}
\end{figure}

In Fig. \ref{fig1} we show the logarithm of the density of states of the 2D 
Ising model on a square lattice with nearest neighbors interactions obtained by 
Wang-Landau sampling for two lattice 
sizes: $L=32,64$. One can see that indeed the slopes are not infinite at the 
ground state. Moreover, this behavior is also observed in the exact results 
obtained by Beale \cite{Beale96}. This situation resembles that of the 
magnetization in the Ising model, which for an infinite system drops to zero 
exactly at the critical temperature but presents a tail for finite-size 
lattices.

For any finite-size lattice of a model with a discrete energy landscape, the 
smallest $\varDelta E$ from the ground state is constant for 
all lattice sizes:
\begin{equation}\label{delta_e}
\varDelta E=const.
\end{equation}
Therefore, the limit in the derivative
\begin{equation}\label{derivative}
\frac{\partial S}{\partial E}=\lim_{\varDelta E \to 0}\frac{\varDelta S}{\varDelta E}
\end{equation}
becomes exact only if $L\rightarrow\infty$ ($E_{min}\rightarrow-\infty$), where 
$L$ is the linear lattice size. As a result, what we have 
to do is to investigate how $\frac{\varDelta S}{\varDelta E}$ diverges as 
$L\rightarrow\infty$. Let us consider three two-dimensional systems:
the Ising model, the $q=3$ Potts model and the $q=4$ Potts model on square
lattices and with nearest neighbors interactions. 

The 2D Ising model for a ferromagnet on a square lattice is described by the 
Hamiltonian
\cite{barkema}
\begin{equation}\label{ising}
{\cal H} = -J \sum_{\langle i,j \rangle} \sigma_i  \sigma_j ,
\end{equation}
where $\sigma_i=\pm1$ represents a spin at lattice site $i$ and the 
notation $\langle i,j \rangle$ indicates that the sites $i$ and $j$ appearing 
in the sum 
are nearest neighbors on the lattice with interaction constant $J>0$. In the 
ground state all the spins are aligned in the same direction, $\sigma_i=+1$, 
for instance. If a single spin is flipped, it looses four 
negative links and gains four positive ones. The energy gap between the ground
state and the next state is therefore $\varDelta E=4-(-4)=8$, where we
take the coupling constant $J$ as 1.

By its turn, the Potts model, proposed by Potts in the early 1950's is an 
extension
of the two states Ising model to $q>2$ states. In this model, to each  lattice 
site is attached a spin variable $\sigma_i$ (defined on each site $i$) which
takes on integer values $1, \ldots, q$. Adjacent sites have an attractive
interaction energy $-J$ whenever they are equal or $0$ otherwise. The
Hamiltonian of the $q$-states ferromagnetic model ($J>0$) on a square 
lattice can be written as
 \cite{barkema}
\begin{equation}
  {\cal H} = -J \sum_{\langle i,j \rangle} \delta_{\sigma_i  \sigma_j},
\end{equation}
where $\delta$ is the Kronecker $\delta-$symbol, and the sum runs over all
nearest neighbors of $\sigma_i$. The ground state has $q$ distinct 
configurations with all spins in the state $1,2,...$ or $q$. If a single spin 
is changed it looses four negative links and does not gain any link, since all 
the neighbors will have a different label. In this case the difference between
the state of minimum energy and the next one is $\varDelta E=0-(-4)=4$, where 
again we set $J=1$.

In the Ising model the ground state has two configurations and the next 
higher level $2L^2$, yielding $\varDelta S=\ln2+2\ln L-\ln2$ and
$\varDelta E=8$, giving
\begin{equation}\label{derivative_i2}
\frac{\varDelta S}{\varDelta E}=\frac{\ln L}{4}=0.25\ln L.
\end{equation}
For notational simplicity we set $k_B=1$.

The $q=3$ Potts model has three configurations in the ground state and $6L^2$ 
in the first higher level. In this case we have $\varDelta
S=\ln6+2\ln L-\ln3=\ln2+2\ln L$ and $\varDelta E=4$, such that
\begin{equation}\label{derivative_p3}
\frac{\varDelta S}{\varDelta E}=\frac{\ln 2}{4}+\frac{\ln L}{2}=0.5\ln(2^{1/2}L)
\end{equation}

Finally for the $q=4$ Potts model we have four configurations in the ground 
state and $12L^2$ in the second level and here $\varDelta
S=\ln12+2\ln L-\ln4=\ln3+2\ln L$ and $\varDelta E=4$, resulting in
\begin{equation}\label{derivative_p4a}
\frac{\varDelta S}{\varDelta E}=\frac{\ln 3}{4}+\frac{\ln L}{2}=0.5\ln(3^{1/2}L)
\end{equation}

In all examples above we see that $\frac{\varDelta S}{\varDelta E}$ has a 
logarithmic dependence on $L$ given by
\begin{equation}\label{derivative_p4b}
\frac{\varDelta S}{\varDelta E}=a\ln(bL).
\end{equation}

Each model has a couple of parameters $a$ and $b$, which governs the 
way $\frac{\varDelta S}{\varDelta E}$ diverges when
$L\rightarrow\infty$. For these models we have therefore calculated 
the exact values to $a$ and $b$, which we display in Table
\ref{table1}.
\begin{table}[!hbt]
\begin{center}
\begin{tabular}{p{0.1cm} p{2.0cm} p{1.0cm} p{1.0cm}  }
\hline\hline
& Model         & $a$     & $b$  \\ \hline
& Ising         & $0.25$  & $1$  \\
& $q=3$ Potts   & $0.5$   & $\sqrt{2}$ \\
& $q=4$ Potts   & $0.5$   & $\sqrt{3}$ \\
\hline\hline
\end{tabular}\
\caption{Exact values of the parameters $a$ and $b$ for the Ising model, the 
$q=3$ Potts model and the $q=4$ Potts model.}
\label{table1}
\end{center}
\end{table}

\section{Entropic Sampling Simulations}
\label{sec:2}

The conventional Wang-Landau method \cite{WangLandau01a,WangLandau01b} is based 
on the fact that if one performs a random walk in energy 
space with a probability proportional to the reciprocal of the density of 
states, a flat histogram is generated for the energy distribution. Since the 
density of states produces huge numbers, instead of estimating $\Omega (E)$, the 
simulation is performed for $S(E)\equiv\ln \Omega(E)$. At
the beginning, we set $S(E)=0$ for all energy levels. The random walk in 
the energy space runs through all energy levels from $E_{min}$ to
$E_{max}$ with a probability $p(E\rightarrow E')=\min(\exp{[S(E)-S(E')]},1)$, 
where $E$ and $E'$ are the energies of the current and the new possible 
configurations, respectively. In fact we begin with any configuration 
(a ground state configuration, for example), and a new possible configuration 
is obtained by changing a single spin state. If the current density of states 
of this energy level $E'$ is less than or equal to that of the present energy 
level $E$, then the configuration is accepted, otherwise we take a random number 
$r$, such that $0<r<1$ and accept this new configuration if 
$r<\Omega(E)/\Omega(E')$ (or $\ln \Omega(E)-\ln \Omega(E')>\ln r$, since we 
are simulating $\ln \Omega(E)$). Then, for this new accepted level or for the 
previous one, we update the histogram $H(E)\rightarrow H(E)+1$ and 
$S(E)\rightarrow S(E)+F_{i}$, where $F_{i}=\ln f_{i}$, with $f_{0}\equiv 
e=2.71828...$ and $f_{i+1}=\sqrt{f_{i}}$ ($f_{i}$ is the so-called modification 
factor). The flatness of the histogram is checked after a certain number of 
Monte Carlo steps (MCS) and usually the histogram is considered flat if 
$H(E)>0.8\langle H \rangle$, for all energies, where $\langle H \rangle$ is an 
average over energies. If the flatness condition is fulfilled, we update the 
modification factor to a finer one and reset the histogram $H(E)=0$. The 
entropic algorithm described above may be formally obtained from the Metropolis 
algorithm if one replaces the Boltzmann's factor $e^{-E/k_BT}$ by the reciprocal 
of the density of states $1/\Omega(E)$. A histogram constructed during the 
Metropolis simulations is given by $H(E)=\Omega(E)e^{-E/k_BT}$ and is similar to 
a Gaussian distribution. If the Boltzmann's factor would be replaced by the 
reciprocal of the density of states, the resulting histogram would be a 
constant. This is the reason why the flatness criterion is used during the 
Wang-Landau simulations to update the modification factor. Ideally, all the 
energy levels should be equally visited during the simulations. 

Recent works \cite{Caparica12,Caparica14,Ferreira12a,Ferreira12b} have 
demonstrated that (a) instead of updating the density of states after every 
move, one ought to update it after each Monte Carlo sweep \cite{MCS}(this 
providence avoids taking into account highly correlated configurations when 
constructing the density of states); (b) WL sampling should be carried out only 
up to $\ln f=\ln f_{final}$ defined by the canonical averages during the 
simulations (this saves CPU time, discarding unnecessary long simulations); and 
(c) the microcanonical averages should not be accumulated before $\ln f \leq ln 
f_{micro}$ defined by a previous study of the microcanonical averaging during 
the simulation (the ruled out WL levels in these averages correspond to a 
microcanonical termalization, since the initial configurations do not match 
those
of maximum entropy). The adoption of these easily implementable changes leads 
to more accurate results and saves computational time. They investigated the 
behavior of the maxima of the specific heat
\begin{equation}\label{cv}
 C(T)=\frac{\langle(E-\langle E\rangle)^2\rangle}{T^2}
\end{equation}
and the susceptibility
\begin{equation}\label{ki}
 \chi(T)=L^2\frac{\langle(m-\langle m\rangle)^2\rangle}{T},
\end{equation}
where $E$ is the energy of the configurations and $m$ is the corresponding 
magnetization per spin, during the WL sampling for the Ising model on a square 
lattice.  They observed that a considerable part of the conventional Wang-Landau 
simulation is not very useful because the error saturates. They demonstrated in 
detail that in general no single simulation run converges to the true value, but 
to a particular value of a Gaussian distribution of results around the correct 
value. The saturation of the error coincides with the convergence to this value. 
Continuing the simulations beyond this limit leads to irrelevant variations in 
the canonical averages of all thermodynamic variables. A later work 
\cite{Caparica14} proposes a criterion for halting the simulations. Applying WL 
sampling to a given model, beginning from $f_{5}$, we calculate the temperature 
of the peak of the specific heat defined in Eq. \eqref{cv} using the current 
$\Omega (E)$ and from this time forth this mean value is updated whenever the 
histogram is checked for flatness. When the histogram is considered flat, we 
save the value of the temperature of the peak of the specific heat $T_c(0)$. We 
then update the modification factor $f_{i+1}=\sqrt{f_{i}}$ and reset the 
histogram $H(E)=0$. During the simulations with this new modification factor we 
continue calculating the temperature of the peak of the specific heat $T_c(t)$ 
whenever we check the histogram for flatness and we also calculate the following 
checking parameter
\begin{equation}\label{eps}
 \varepsilon=|T_c(t)-T_c(0)|.
\end{equation}

\begin{table*}[!ht]
\centering
\begin{tabular}{ l c l c c l c c}
\multicolumn{2}{c}{Ising model} & & \multicolumn{2}{c}{$q=3$ Potts model} & &\multicolumn{2}{c}{$q=4$ Potts model} \\
\cline{1-2} \cline{4-5} \cline{7-8}
       \multicolumn{1}{c}{$a$}     &    $b$    & &    $a$     &    $b$    & &    $a$      & $b$       \\
\cline{1-2} \cline{4-5} \cline{7-8}
0.24949(39) & 1.0063(58)  & & 0.49968(84) & 1.4202(95) & & 0.50068(77) & 1.721(11)\\
0.25036(47) & 0.9953(69)  & & 0.50059(65) & 1.4061(73) & & 0.49965(83) & 1.736(12)\\
0.25028(24) & 0.9963(36)  & & 0.49978(66) & 1.4159(74) & & 0.49868(71) & 1.754(10)\\
0.24961(29) & 1.0054(43)  & & 0.50055(76) & 1.4077(84) & & 0.4996(11)  & 1.739(16)\\
0.25021(44) & 0.9975(63)  & & 0.50116(85) & 1.4034(95) & & 0.5007(11)  & 1.722(15)\\
0.24972(26) & 1.0056(38)  & & 0.50142(68) & 1.3993(75) & & 0.50008(67) & 1.729(10)\\
0.25028(43) & 0.9957(60)  & & 0.49994(85) & 1.4161(94) & & 0.50012(84) & 1.730(12)\\
0.24991(31) & 1.0026(46)  & & 0.49869(85) & 1.4291(97) & & 0.49853(73) & 1.755(11)\\
0.25038(38) & 0.9957(55)  & & 0.50006(62) & 1.4129(69) & & 0.49967(94) & 1.736(13)\\
0.24956(46) & 1.0054(66)  & & 0.50089(73) & 1.4057(81) & & 0.50153(75) & 1.711(10)\\
\cline{1-2} \cline{4-5} \cline{7-8}
0.24998(11) & 1.0006(15)  & & 0.50028(26) & 1.4116(28) & & 0.49992(29) & 1.733(44)\\
\end{tabular}
\caption{Estimates by entropic sampling simulations of the parameters $a$ and $b$ for the Ising model,
the $q=3$ Potts model and the $q=4$ Potts model.}
\label{table2}
\end{table*}
If the number of MCS before verifying the histogram for flatness is 
chosen not too large, say $10,000$, then during the simulations with the same 
modification factor the checking parameter $\varepsilon$ is calculated many 
times. If $\varepsilon$ remains less than $10^{-4}$ until the histogram meets 
the flatness criterion for this WL level, then we save the density of states and 
the microcanonical averages and stop the simulations. When one adopts this 
criterion for halting the simulations, different runs stop at different final 
modification factors. It was also observed in Ref. \cite{Caparica14} that two 
independent similar finite-size scaling procedures can lead to very different 
results for the critical temperature and exponents, which often do not agree 
within the error bars. The way to overcome this difficulty is to carry out 10 
independent sets of finite-size scaling simulations. In the present work, for 
each of theses sets and for each model (Ising, $q=3$ and $q=4$ Potts models), we 
performed simulations for $L=8,12,16,20,24,28,32,36,40,44,48,52,56,64,72$, and 
$80$ with $n=32,32,28,28,28,24,24,24,20,20,20,16,16,16,\\ 12$, and $12$ 
independent runs for each size, 
respectively.

\section{Simulational Results}
\label{sec:3}

In fact, we used the outcomes of the simulations described in
Ref. \cite{Caparica14} with $\varepsilon<10^{-4}$ for the Ising model and Ref. 
\cite{Caparica15} for the $q=3$ and $q=4$ Potts models. The final resulting 
values for $\frac{\varDelta S}{\varDelta E}$ in each case were obtained as an 
average over all sets. In Table \ref{table2} we display the values obtained in 
our entropic simulations for $a$ and $b$ for each of the considered models. The 
ten initial lines correspond to results obtained in each set and the last line 
is an average over all sets. In carrying out theses averages we tried two 
alternative ways: an average with unequal uncertainties \cite{Wong97} and a 
direct average neglecting the error bars. We observed that the second procedure 
leads to better results with more reliable error bars and these are the results 
we display in the last line of the Table \ref{table2}. One can see that our 
results for $a$ and $b$ agree within error bars equal to $\pm1\sigma$ with the 
exact results shown in Table \ref{table1} in all cases, reaching the very limit 
only for the parameter $a$ of the $q=3$ Potts model.
\begin{figure}[!th]\centering
\begin{center}
 \includegraphics[width=.7\linewidth,angle=-90]{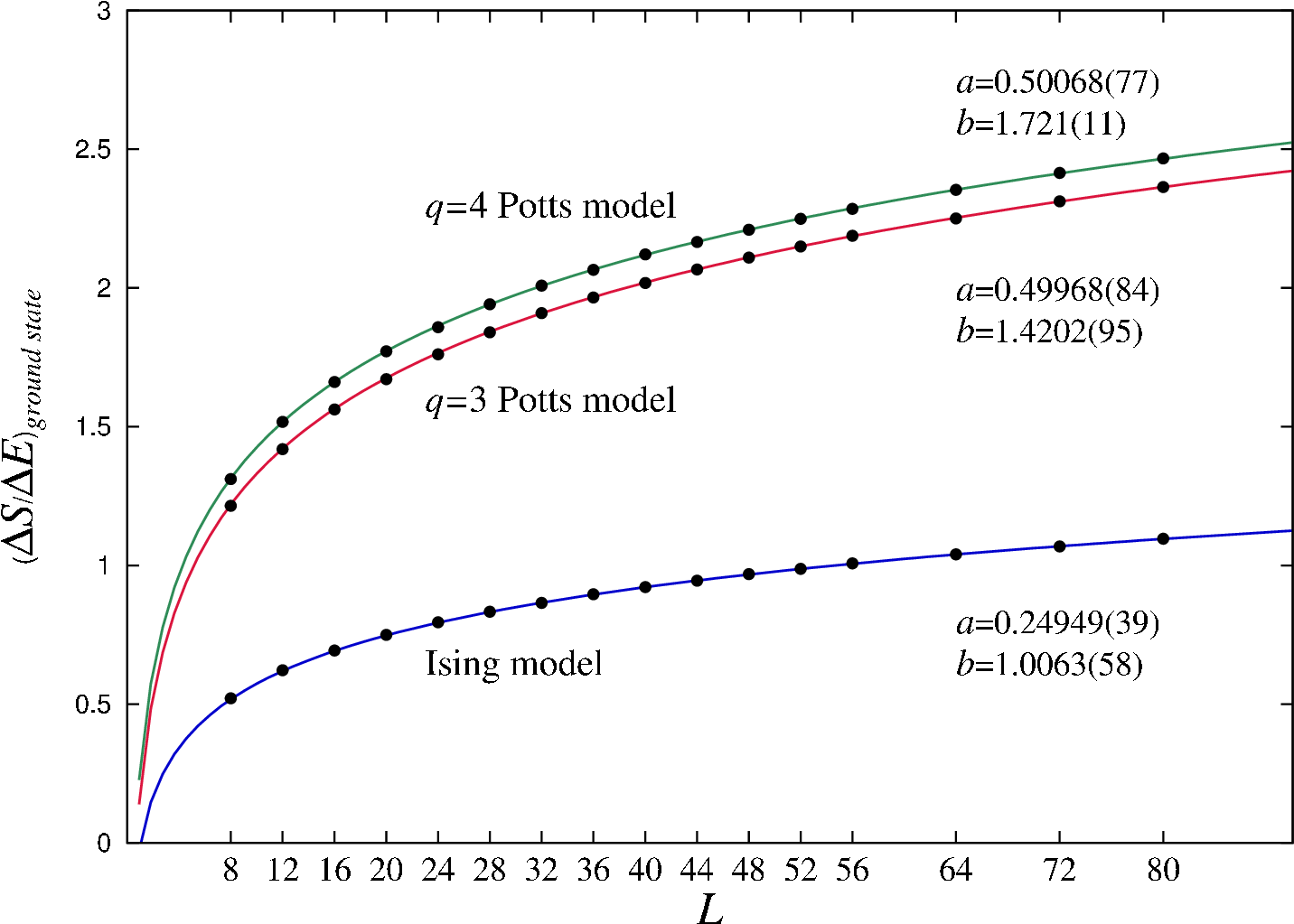}
\end{center}
\caption{(Color online) Size dependence of $\frac{\varDelta S}{\varDelta E}$ 
at the ground state of the first set of simulations. The lines are best fitting 
curves of these data to $a\ln(bL)$. The error bars are smaller than the 
symbols.}
\label{fig2}
\end{figure}

In Fig. \ref{fig2} we show the dependence of $\frac{\varDelta S}{\varDelta E}$ 
at the ground state on the lattice sizes using the outcomes of the first set of 
simulations, along with the best fitting curves to $a\ln(bL)$. The agreement is 
excellent.

It is noteworthy that this couple of parameters $a$ and $b$ represents a new 
interesting issue of investigation in studies using entropic sampling. A 
possible challenge would be to estimate them in the case of polymers due to the 
difficulty of achieving good statistics near the ground state. Furthermore it is 
an open question if all systems have the logarithmic behavior or eventually it 
could arise, for example, a power law with a coefficient and an exponent: 
$KL^\kappa$.

\section{Conclusions}
\label{sec:4}

To summarize, in this work we analyzed the behavior of the ground state 
microcanonical temperature in entropic sampling simulations of spin lattice 
models. We verified that the expected divergence not observed in most studies 
are indeed related to finite-size effects. We verified that $\frac{\varDelta 
S}{\varDelta E}$ in the ground state diverges logarithmically and we proposed an 
analytic expression $a\ln(bL)$ to fit this behavior. Our exact and numerical 
results exhibit an excellent agreement.  We have also shown that it is 
straightforward to calculate the constant parameters $a$ and $b$ related to the 
logarithmic behavior of the entropy at the ground state for the considered 
models.  A further analysis would be to verify if this analytic expression still 
holds for systems with continuous energy spectrum.

\begin{acknowledgements}
We acknowledge the computer resources provided by LCC-UFG.
\end{acknowledgements}
\vfill

\end{document}